\documentclass[12pt]{article}

\usepackage{latexsym,amsmath,amscd,amssymb,graphics,cite}
\usepackage{enumerate}

\usepackage{graphicx}
\usepackage{framed}
\usepackage[colorlinks]{hyperref}
\usepackage{url}
\usepackage{cite}
\usepackage{mathrsfs}

\usepackage[all]{xy}

\makeatletter

\@addtoreset{figure}{section}
\def\thefigure{\thesection.\@arabic\c@figure}
\def\fps@figure{h, t}
\@addtoreset{table}{bsection}
\def\thetable{\thesection.\@arabic\c@table}
\def\fps@table{h, t}
\@addtoreset{equation}{section}

\makeatother

\newtheorem{theorem}{Theorem}

\newtheorem{remark}[theorem]{Remark}

\numberwithin{theorem}{section}

\def\be{\begin{equation}}
\def\ee{\end{equation}}
\def\bea{\begin{eqnarray}}
\def\eea{\end{eqnarray}}
\def\ba{\begin{array}}
\def\ea{\end{array}}

\def\bOm{\boldsymbol{\Omega}}

\let\<\langle
\let \>\rangle

\newcommand{\rem}[1]{}

\newcommand{\bGam}{\boldsymbol{\Gamma}}

\newcommand{\bN}{\mathbf{N}}

\newcommand{\bchi}{\boldsymbol{\chi}}

\newcommand{\pp}[2]{\frac{\partial #1}{\partial #2}}

\newcommand{\mso}{\mathfrak{so}}

\title{Geometric analysis of noisy perturbations to nonholonomic constraints}
\author{Fran\c{c}ois Gay-Balmaz$^1$ and 
Vakhtang Putkaradze$^2$ \vspace{0.2cm}\\
\small $^1$ LMD/IPSL, CNRS, Ecole normale sup\'erieure, PSL Research University,\\
\small Ecole polytechnique, Universit\'e Paris-Saclay, Sorbonne Universit\'es,\\
\small UPMC Univ Paris 06, 24 rue Lhomond, 75005, Paris, France.\\
\small $^2$Department of Mathematical and Statistical Sciences\\
\small University of Alberta, Edmonton, AB   T6G 2G1 Canada
\\
}

\begin{document}
\maketitle

\begin{abstract} We propose two types of stochastic extensions of nonholonomic constraints for mechanical systems. Our approach relies on a stochastic extension of the Lagrange-d'Alembert framework. We consider in details the case of invariant nonholonomic systems on the group of rotations and on the special Euclidean group. Based on this, we then develop two types of stochastic deformations of the Suslov problem and study the possibility of extending to the stochastic case the preservation of some of its integrals of motion such as the Kharlamova or Clebsch-Tisserand integrals.
\end{abstract}

\section{Introduction} 
 
Nonholonomic constraints have been an important part of classical mechanics for few centuries, see, e.g., \cite{Bloch2003} and\cite{ArKoNe1997}. These types of constraints, incorporating the velocity in an essential way, usually arise  as an appropriate mathematical idealization of some contact conditions between parts of a mechanical system. For example, the no-slip rolling conditions are in general nonholonomic except for the simplest cases. The nonholonomic rolling condition states that the infinitesimally small point of contact experiences enough total force not to move with respect to the substrate. In other words, the force per area for the rolling constraint must be infinite for the nonholonomic condition to be valid. 

Since nonholonomic constraints are mathematical idealizations, there has been considerable interest in deriving nonholonomic systems as a limit of physically realizable physical forces. As an example, we shall point to the monograph \cite{ArKoNe1997} where the nonholonomic equations are discussed from the point of view of limits of viscous forces and parameter limits. 
Alternatively, realising the nonholonomic constraint as an appropriate limit of potential forces can be also useful for example for the quantization of nonholonomic systems, as discussed in \cite{BlRo2008}.

In these notes, we consider the effects of introducing noise in the nonholonomic constraints directly,  i.e., study the case of \emph{stochastic constraints}. Physically our study explores what would happen if there is a stochastic error in the nonholonomic constraint. For example, a ball or a disk rolling on a surface may have random slippage introducing a mismatch in velocity. Alternatively, that surface may have a small roughness or exhibits microscopic random motions forcing the constraint to have a stochastic part. These cases are analysed and discussed in details in the recent paper by the authors \cite{FGBPu2016}. The main focus of that paper was the consideration of energy conservation, as well as other integrals of motion, under stochastic constraints.  The reader is encouraged to consult that paper for the in-depth consideration of the rolling ball case, as well as other particular applications. The present notes should be viewed as supplementary to that paper as we develop in details two types of stochastic deformations of the Suslov problem in Section \ref{sec:Suslov} and study the possibility of extending to the stochastic case the preservation of some integrals of motion such as the Kharlamova or Clebsch-Tisserand integrals, see \cite{Ko1985}.

Stochastic perturbations of Hamiltonian systems have been considered before in 
\cite{Ne1967,Bi1981,LaOr2008,BROw2009,Ho2015,GBHo2017}. Closer to our interests here, 
the stochastic extensions of nonholonomically constrained systems, obeying \emph{exact} nonholonomic constraints, was studied in \cite{Ho2010,HoRa2015}. In that work, the focus was on physical systems that are sensitive to noise, while preserving the nature of the nonholonomic constraints exactly. Our paper considers, in some sense, the opposite problem of intrinsically (and somewhat loosely speaking) non-noisy system with noisy constraints. Of course, the stochasticity of constraints leads to stochasticity of the system itself, so the full system is stochastic. However, the stochasticity is quite different from the one discussed before, as it allows for the preservation of integrals of motion. For example, the energy can be conserved under quite general perturbations of the constraints, as we derive in the general form below in Section \ref{sec:stochastic_constraints}.

\section{The Lagrange-d'Alembert equations in nonholonomic mechanics}
\label{sec:LdA}

In this section, we quickly recall the Lagrange-d'Alembert principle in nonholonomic mechanics and two important classes of nonholonomic systems on Lie groups with symmetry.

\subsection{Lagrange-d'Alembert principle}

Consider a mechanical system with configuration manifold $Q$ and Lagrangian $L= L( q , \dot q): TQ \rightarrow \mathbb{R}  $ defined on the tangent bundle $TQ$ of the manifold $Q$. Let us assume that the mechanical system is subject to a linear constraint on velocity, encoded in a nonintegrable distribution $ \Delta \subset TQ$ on $Q$. The equations of motion for the nonholonomic system can be derived from the \textit{Lagrange-d'Alembert principle}
\begin{equation}\label{LdA} 
\left.\frac{d}{d\varepsilon}\right|_{\varepsilon=0} \int_{0 }^{ T}L(q _\varepsilon (t) , \dot q_ \varepsilon (t) )dt=0,
\end{equation}
where $\dot q(t) \in\Delta $ and for variations $\delta q (t) = \left.\frac{d}{d\varepsilon}\right|_{\varepsilon=0} q _\varepsilon(t)$ 
of the curve $q(t)$ such that $ \delta q (t)  \in \Delta $ and $ \delta q(0)= \delta q(T)=0$. Note that the curves $q_\varepsilon (t)$ do not need to satisfy the constraint when $ \varepsilon \neq 0$.

From \eqref{LdA} one derives the Lagrange-d'Alembert equations 
\begin{equation}\label{LdA_equations}
\frac{d}{dt} \frac{\partial L}{\partial \dot q}- \frac{\partial L}{\partial q}\in  \Delta(q)  ^\circ,  \quad \dot q\in \Delta (q),
\end{equation}
where $ \Delta (q) \subset T_qQ$ is the vector fiber of $ \Delta$ at $q \in Q$ and $ \Delta (q) ^\circ \subset T^*_qQ$ is the annihilator of $ \Delta (q)$ in $T^*_q Q$. When there are no constraints, \eqref{LdA_equations} reduces to the Euler-Lagrange equations for the unconstrained system.

\medskip

It is also useful to write \eqref{LdA_equations} explicitly in local coordinates as follows. 
Let $ \omega \in \Omega ^1 (Q, \mathbb{R}  ^m )$ be a $ \mathbb{R} ^m $-valued one-form such that $ \Delta (q)=\ker ( \omega (q))$. We assume that the $m$ components of $\omega $ are independent. One can choose, in a neighborhood of each point, a local coordinate chart such that the one-form $ \omega $ reads
\begin{equation}\label{local_expression} 
\omega ^a (q)= ds ^a + A_ \alpha ^a (r,s)dr ^\alpha , \quad a=1,...,m,
\end{equation}
where $q=(r,s) \in U \subset \mathbb{R}  ^{n-m} \times \mathbb{R}  ^m $. In these coordinates, equations \eqref{LdA_equations} read
\begin{equation} 
\label{LdA_local}
\frac{d}{dt} \frac{\partial L}{\partial \dot r ^\alpha }- \frac{\partial L}{\partial r ^\alpha }= A _\alpha ^a \left( \frac{d}{dt} \frac{\partial L}{\partial \dot s ^a }- \frac{\partial L}{\partial s ^a  }\right) , \;\; \alpha =1,...,n-m,\quad \dot s ^a =- A _\alpha ^a \dot r ^\alpha ,\;\; a=1,...,m.
\end{equation} 
For later comparisons with the case of stochastic nonholonomic constraints, it is instructive to rewrite the Lagrange-d'Alembert equations in terms of the \textit{constrained Lagrangian} 
\[
L_C( r ^\alpha , s ^a , \dot r ^\alpha ):= L(  r ^\alpha , s ^a , \dot r ^\alpha , -A_ \alpha ^a (r,s)\dot r ^\alpha ),
\]
which is obtained by substituting the expression for $\dot s^a$ from the constraint in \eqref{LdA_local} into the Lagrangian $L(r,\dot r, s, \dot s)$.
We get
\begin{equation}\label{constrained_L} 
\frac{d}{dt}\frac{\partial L_C}{\partial \dot r ^\alpha }- \frac{\partial L_C}{\partial r ^\alpha }+ A _\alpha ^a \frac{\partial L_C}{\partial s ^a }=- \frac{\partial L}{\partial \dot s ^b}  B ^b _{ \alpha \beta } \dot r^\beta,
\end{equation} 
where we have defined
\begin{equation}\label{def_B} 
B ^b _{\alpha \beta } =\frac{\partial A ^b _ \alpha }{\partial r ^\beta }-\frac{\partial A ^b _ \beta  }{\partial r ^ \alpha  }+ A ^a _\alpha \frac{\partial A ^b _ \beta  }{\partial s ^a  }- A_\beta ^\alpha \frac{\partial A ^b _ \alpha }{\partial s ^ \alpha }. 
\end{equation} 
We refer to \cite{Bloch2003} and \cite{BlKrMaMy1996}  for a detailed treatment of the Lagrange-d'Alembert equations.

\subsection{Invariant nonholonomic systems on Lie groups}

Let us assume that the configuration manifold is a Lie group $Q=G$ and that the Lagrangian $L:TG \rightarrow \mathbb{R}  $ is (left) $G$-invariant. We can thus define the associated reduced Lagrangian $\ell: \mathfrak{g}  \rightarrow \mathbb{R}  $ on the Lie algebra $ \mathfrak{g}$ of $G$, such that $L(g,v)=\ell( g ^{-1} v)$, for all $g \in G$ and $v \in T_gG$.

We also assume that the distribution $ \Delta  \subset TG$ is (left) $G$-invariant. In terms of the one-form $\omega  \in \Omega ^1 (G, \mathbb{R}  ^k )$, we thus have $\omega  (g) \cdot v=  \omega (e)\cdot g^{-1} v$, for all $(g,v) \in TG$, and we define the subspace
\[
\mathfrak{g}  ^\Delta := \Delta (e)=\{ \xi \in \mathfrak{g}  \mid \omega (e) \cdot \xi =0\} \subset \mathfrak{g}.
\]
The Lagrange-d'Alembert equations for $L$ and $ \Delta $ can be equivalently written in terms of $ \ell$ and $ \mathfrak{g}  ^ \Delta $ as
\begin{equation}\label{EPS} 
\frac{d}{dt} \frac{\delta \ell}{\delta \xi }- \operatorname{ad}^*_ \xi \frac{\delta \ell}{\delta \xi } \in (\mathfrak{g}  ^\Delta ) ^\circ, \qquad \xi \in\mathfrak{g}  ^\Delta.
\end{equation} 
These are the \textit{Euler-Poincar\'e-Suslov equations}, see \cite{Bloch2003,Kozlov1988}. In \eqref{EPS}, $ (\mathfrak{g}  ^\Delta ) ^\circ :=\{ \mu \in \mathfrak{g}  ^\ast \mid \left\langle \mu , \xi \right\rangle =0, \;\text{for all} \; \xi \in \mathfrak{g}  ^\Delta\} \subset \mathfrak{g}  ^\ast $ denotes the annihilator of $\mathfrak{g}  ^\Delta $ in $\mathfrak{g}  ^\ast $, $\frac{\delta \ell}{\delta \xi } \in\mathfrak{g}  ^\ast $ defined by
\[
\left\langle \frac{\delta \ell}{\delta \xi }, \eta \right\rangle = \left.\frac{d}{d\varepsilon}\right|_{\varepsilon=0} \ell( \xi + \varepsilon\eta ),
\]
is the functional derivative of $ \ell$, and $ \operatorname{ad}^*_ \xi :\mathfrak{g}  ^\ast \rightarrow\mathfrak{g} ^\ast $, defined by $ \left\langle\operatorname{ad}^*_ \xi \mu ,\eta \right\rangle =\left\langle\mu , [\xi  , \eta ]\right\rangle $,  is the coadjoint operator.

Let us illustrate these concepts for the case of the Lie group $G=SO(3)$ of $3 \times 3$ rotation matrices. Physically, such a group defines the attitude of a rigid body about a fixed point in space, usually assumed to be the center of mass. The Lie algebra of this group, denoted $\mso(3)$, is the algebra of all skew-symmetric matrices, identified with $\mathbb{R}^3$ through the hat map $\;\widehat{\;}:\mathbb{R}  ^3\rightarrow \mso(3)$:
\begin{equation} 
\widehat{a}_{i j} = - \epsilon_{ijk} a_k , \quad \mathbf{a}=\left(a_1, a_2, a_3 \right)^T \in \mathbb{R}^3 \, . 
\label{so3def} 
\end{equation} 
For this definition, the Lie algebra bracket is related to the cross product as
\begin{equation} 
\left[ \widehat{ \mathbf{a} } , \widehat{ \mathbf{b} } \right]  = \widehat{ \mathbf{a} \times \mathbf{b} } \, , \quad \forall \, \,
\widehat{ \mathbf{a} } , \widehat{ \mathbf{b} } \in \mso(3) .
\label{widehatproperties}
\end{equation} 
Let us now consider the equations of motion for a left-invariant Lagrangian $L:TSO(3) \rightarrow \mathbb{R} $, with associated reduced Lagrangian $\ell: \mso(3) \rightarrow \mathbb{R}  $. The reduced velocity (the variable $ \xi $ above) has the physical meaning of the body angular velocity $\widehat{\boldsymbol{\Omega} }=\Lambda^T \dot \Lambda \in \mso(3)$, where $ \Lambda (t) \in SO(3)$ is the attitude of the body. The constraint reads $\bOm \cdot \mathbf{a} =0$ for some constant vector $ \mathbf{a}  \in \mathbb{R}^3$. This is the formulation of the classical Suslov problem \cite{Suslov1951}, which is perhaps one of the simplest nonholonomic problems having a non-trivial behavior. The equations of motion \eqref{EPS} are written explicitly as 
\begin{equation} 
\frac{d }{d t} \frac{ \delta \ell}{ \delta \bOm}+ \bOm \times \frac{ \delta \ell}{ \delta \bOm}= \lambda \mathbf{a}  \, , \quad \bOm \cdot \mathbf{a}  =0 ,
\label{Suslov_simple}
\end{equation} 
where $\lambda$ is a scalar enforcing the constraint. We will study Suslov's problem in considerable more details, and in a more general formulation,  in Section \ref{sec:Sus_prob} below. 
\color{black}

\subsection{Rolling ball type systems on semidirect products}\label{geom_setting} 

Let us consider a slightly more involved situation, namely, the case of a (possibly nonhomegeneous) rolling ball on a flat surface.
In this case the configuration Lie group is given by the semidirect product
\[
SE(3)= SO(3) \,\circledS\, \mathbb{R}  ^3 \ni ( \Lambda , \mathbf{x} ),
\]
with group multiplication $( \Lambda_1  , \mathbf{x} _1 )(\Lambda _2 , \mathbf{x} _2 )= ( \Lambda _1\Lambda _2 , \mathbf{x} _1 + \Lambda _1 \mathbf{x} _2 )$.
Here $ \Lambda $ describes the attitude of the ball and $ \mathbf{x} $ the position of its center of mass. As opposed to the previous case, the Lagrangian $L:TSE(3) \rightarrow \mathbb{R}  $ and the distribution $ \Delta \subset TSE(3)$ are only invariant under a subgroup of $SE(3)$, namely, the group $S ^1 \,\circledS\, \mathbb{R}  ^3$, where $S^1 $ is the group of rotations around the $ \mathbf{z} $-axis. The quotiented velocity space $TSE(3)/ (S ^1 \,\circledS\, \mathbb{R}  ^3)$ is identified as follows
\[
TSE(3)/ (S ^1 \,\circledS\, \mathbb{R}  ^3)\simeq \mathfrak{se}(3) \times S ^2\ni ( \widehat{\boldsymbol{\Omega} }, \mathbf{Y} ,\boldsymbol{\Gamma} ), 
\]
where $\widehat{\boldsymbol{\Omega} }= \Lambda ^T \dot \Lambda $ is the body angular velocity, $ \mathbf{Y} = \Lambda ^{-1} \dot{ \mathbf{x} }$ is the body velocity of the center of mass, and $ \boldsymbol{\Gamma}= \Lambda ^{-1} \mathbf{z} $ is (minus) the direction of gravity measured from the body frame.

The reduced Lagrangian $\ell$ associated to $L$ is now defined as $\ell: \mathfrak{se}(3) \times S ^2 \rightarrow \mathbb{R}  $. The reduced constraint associated to $ \Delta \subset TSE(3)$ is a $ \boldsymbol{\Gamma} $-dependent subspace $ \mathfrak{se}(3)^ \Delta ( \boldsymbol{\Gamma} ) \subset  \mathfrak{se}(3)$, $\boldsymbol{\Gamma} \in S ^2$. We shall assume the form
\begin{equation}\label{constraint} 
\mathfrak{se}(3)^ \Delta ( \boldsymbol{\Gamma} )=\{( \boldsymbol{\Omega} , \mathbf{Y} ) \in \mathfrak{se}(3) \mid \mathbf{Y} = \alpha ( \boldsymbol{\Gamma} ) \cdot\boldsymbol{\Omega} \},
\end{equation} 
where $ \alpha (\boldsymbol{\Gamma} ): \mathbb{R} ^3 \rightarrow \mathbb{R}^3 $ is a $\boldsymbol{\Gamma} $-dependent linear map. For an explicit expression of the map $\alpha$ in terms of $\bGam$ and physical parameters of the rolling ball, and appropriate physical discussion of slipping without rolling, we refer the reader to \cite{Bloch2003,Ho2008} and in the stochastic formulation to \cite{FGBPu2016}.

In terms of $L:T SE(3) \rightarrow \mathbb{R}  $ and $ \Delta \subset TSE(3)$, the equations of motion are given by the Lagrange-d'Alembert equations. In terms of the reduced Lagrangian $\ell$ and reduced distribution $ \mathfrak{se}(3)  ^\Delta (\boldsymbol{\Gamma} )$, these equations  take the form
\begin{equation}\label{reduced_equations} 
\left\{
\begin{array}{l}
\vspace{0.2cm}\displaystyle \frac{d}{dt}  \frac{\delta \ell}{\delta \boldsymbol{\Omega}  }   + \boldsymbol{\Omega} \times \frac{\delta \ell}{\delta\boldsymbol{\Omega}  }+ \mathbf{Y}\times  \frac{\delta \ell}{\delta \mathbf{Y} } + \alpha( \boldsymbol{\Gamma} )^{T} \cdot  \left( \frac{d}{dt}  \frac{\delta \ell}{\delta\mathbf{Y} } +\boldsymbol{\Omega} \times  \frac{\delta \ell}{\delta \mathbf{Y} }\right) =\frac{\delta\ell}{\delta\boldsymbol{\Gamma} }\times\boldsymbol{\Gamma} \\
\vspace{0.2cm} \displaystyle\frac{d}{dt}\boldsymbol{\Gamma}  + \boldsymbol{\Omega} \times\boldsymbol{\Gamma}  =0, \quad \mathbf{Y} = \alpha(\boldsymbol{\Gamma} ) \cdot \boldsymbol{\Omega},
\end{array}
\right.
\end{equation} 
where $ \frac{\delta \ell}{\delta \boldsymbol{\Omega} }, \frac{\delta \ell}{\delta \mathbf{Y} }, \frac{\delta \ell}{\delta\boldsymbol{\Gamma} } \in \mathbb{R}  ^3 $ are the partial derivatives of $\ell$.

We refer to \cite{GBYo2015} for the general theory of rolling ball type systems on semidirect products and for a detailed derivation of the reduced Lagrange-d'Alembert equations \eqref{reduced_equations}, both from the point of view of variational principles and from the point of view of Dirac structures.

\section{Stochastic extensions of nonholonomic constraints}\label{Stoch_LdA} 
\label{sec:stochastic_constraints}

Let us fix a configuration manifold $Q$, a Lagrangian function $L: TQ \rightarrow \mathbb{R}  $, and a constraint distribution $ \Delta \subset TQ$. Let us assume that the constraint distribution is written as $ \Delta (q)=\{v \in T_qQ\mid  \omega (q) \cdot  v=0\}$ for a $\mathbb{R}  ^m$-valued one-form $ \omega  \in\Omega^1(Q, \mathbb{R}  ^m )$. We assume that the $m$ components of $\omega $ are independent.

\subsection{Stochastic Lagrange-d'Alembert equations}\label{I_and_II} 

We present two ways to introduce stochasticity in the nonholonomic constraint.

\paragraph{I. Affine stochastic constraint.}
The first way of introducing stochasticity in the constraints consists in replacing the linear constraint $\omega (q) \cdot  v = 0 $ by an affine constraint
\begin{equation}\label{affine}
\omega(q) \cdot v=N,
\end{equation} 
for $N \in \mathbb{R} ^m $.
The stochasticity is inserted in the dynamics through the constraint by assuming that $N(t)$ is a stochastic process verifying the Stratonovich stochastic differential equation
\begin{equation}\label{dN}
\mbox{d}N=F(q,v,N) \mbox{d}t+ \Sigma  (q,v,N) \circ \mbox{d}W,
\end{equation} 
for given functions $F, \Sigma : TQ \times \mathbb{R}  ^m \rightarrow  \mathbb{R}  ^m $.

We associate to this constraint the stochastic Lagrange-d'Alembert equations
\begin{equation}\label{LdA1} 
\left\{
\begin{array}{l}
\vspace{0.2cm}\displaystyle \mbox{d}\frac{\partial L}{\partial v}- \frac{\partial L}{\partial q} \mbox{d}t \in \Delta (q) ^\circ\\
\vspace{0.2cm}\displaystyle \mbox{d}q=v\mbox{d}t\\
\vspace{0.2cm}\displaystyle  \omega (q) \cdot v=N\\
\mbox{d}N= F(q,v, N)\mbox{d}t+ \Sigma (q,v, N) \circ \mbox{d}W,
\end{array}
\right.
\end{equation} 
understood in the Stratonovich sense and whose precise meaning is explained below. In general, the energy of the system, defined by
\begin{equation}\label{energy_L} 
E(q,v):= \left\langle \frac{\partial L}{\partial v},v \right\rangle -L(q,v),
\end{equation} 
is not preserved along the solutions of \eqref{LdA1}.

\paragraph{Stratonovich form.} Taking local coordinates $(r ^\alpha , s ^a )$ such that $ \omega $ takes the expression \eqref{local_expression}, we can rewrite system \eqref{LdA1} as   
\begin{equation}\label{LdA1_rs} 
\left\{
\begin{array}{l}
\vspace{0.2cm}\displaystyle \mbox{d}\frac{\partial L}{\partial u ^\alpha }- \frac{\partial L}{\partial r ^\alpha }\mbox{d}t= A _\alpha ^a (r,s) \left(  \mbox{d}\frac{\partial L}{\partial w ^a }- \frac{\partial L}{\partial s ^a } \mbox{d}t\right) \\
\vspace{0.2cm}\displaystyle \mbox{d}r ^\alpha = u ^\alpha \mbox{d}t, \;\; \mbox{d}s ^a =w ^a \mbox{d}t\\
\vspace{0.2cm}\displaystyle  w ^a + A _\alpha ^a (r,s) u^\alpha = N ^a \\
\mbox{d}N ^a = F ^a (r,s,u,w, N)\mbox{d}t+ \Sigma ^a (r,s,u,w, N) \circ \mbox{d}W.
\end{array}
\right.
\end{equation} 
Defining the constrained Lagrangian as
\begin{equation}\label{def_LC1} 
\mathcal{L} _C(r ^\alpha ,s ^a ,u ^\alpha  , N ^a )= L( r ^\alpha ,s ^a ,u ^\alpha ,N ^a-A _\alpha ^a (r,s) u ^\alpha ),
\end{equation} 
similar computations as in the deterministic case, yield the first equation of \eqref{LdA1_rs} in the form
\begin{equation}\label{LC_form1} 
\mbox{d}\frac{\partial \mathcal{L} _C}{\partial u ^\alpha }- \frac{\partial \mathcal{L} _C}{\partial r ^\alpha }\mbox{d}t+ A _\alpha ^a \frac{\partial\mathcal{L} _C}{\partial s ^a }\mbox{d}t=- \frac{\partial L}{\partial w^b}  \left( B ^b _{ \alpha \beta } w^\beta +\frac{\partial A _\alpha ^b }{\partial s ^a }N ^a \right) \mbox{d}t,
\end{equation} 
where $B$ is defined as in \eqref{def_B}. 
The difference with the deterministic counterpart \eqref{constrained_L}, lies in two facts. Firstly, the constrained Lagrangian $ \mathcal{L} _C$ (and hence its partial derivatives) depends on the stochastic process $N$ through its definition \eqref{def_LC1}. Secondly, there is the additional last term in \eqref{LC_form1} that explicitly involves $N$.

By developing the term $ \mbox{d}\frac{\partial \mathcal{L} _C}{\partial u ^\alpha }$ using the Stratonovich chain rule, and assuming that $ \frac{\partial ^2  \mathcal{L} _C}{\partial u ^\alpha u ^\beta }$ is invertible, we can rewrite the system \eqref{LdA1_rs} in the standard Stratonovich form
for the stochastic processes $(r,s,u,N)$ as
\begin{equation}\label{LdA1_rs_Stratonovich} 
\left\{
\begin{array}{l}
\vspace{0.2cm}\displaystyle \mbox{d}r ^\alpha = u ^\alpha \mbox{d}t, \quad  \mbox{d}s ^a =(N ^a- A _\alpha ^a (r,s) u^\alpha ) \mbox{d}t\\
\vspace{0.2cm}\displaystyle \mbox{d}  u ^\alpha = P(r,s,u, N) \mbox{d}t - \left( \frac{\partial ^2 \mathcal{L} _C}{\partial u ^\alpha \partial u ^\beta } \right) ^{-1} \frac{\partial ^2\mathcal{L} _C}{\partial u ^\beta \partial N ^a } \Sigma _C ^a \circ \mbox{d}W\\
\vspace{0.2cm}\displaystyle \mbox{d}N ^a = F ^a _C \mbox{d}t+ \Sigma ^a_C  \circ \mbox{d}W,
\end{array}
\right.
\end{equation} 
where $F_C$ is defined by $F  _C(r^\alpha , s^a ,u ^\alpha , N ^a ):= F ( r ^\alpha ,s ^a ,u ^\alpha ,N ^a-A _\alpha ^a (r,s) u ^\alpha , N ^a )$, similarly for $ \Sigma _C $, and $P$ is a function depending on the variables through the Lagrangian and its derivatives, as well as the functions $A^a_\alpha$. The function $P(r,s,u, N)$ has an explicit expression involving the partial derivatives of $ \mathcal{L} _C$, the partial derivative $ \frac{\partial L}{\partial w ^b }$, and the quantities $A_ \alpha ^b $, $ \frac{\partial A _\alpha ^b }{\partial s ^a }$, $B_{ \alpha \beta } ^b $.

\rem{ 
\begin{remark}\label{functions_of_noise}{\rm In some examples, like the ones in Section~\ref{sec:shake}, the variable $N$ does not have the meaning of the physical noise experienced by the system. Rather, $N$ is a function of the noise $n$ as well as another variable $q$, i.e., $N= \gamma (q,n)$. Given a Stratonovich SDE $\mbox{d}n= a(q,v,n)\mbox{d}t+ b(q,v,n) \circ \mbox{d}W$ for $n$, one can compute $\mbox{d}N$ by applying the Stratonovich chain rule. To accomplish that, we will need to use the inverted relationship $n=n(N,q)$, assuming this inverse function of $N$ and $q$ exists for $(N,q)$ belonging to the region of interest. For example, for the case of the rolling ball, $\mathbf{N}$ and $\mathbf{n}$ are related by $\mathbf{N}=\Lambda^{-1} \mathbf{n}$, which is invertible for any $\Lambda \in SO(3)$. This will allow us to  deduce the corresponding generalizations of \eqref{LdA1}, \eqref{LdA1_rs}, and \eqref{LdA1_rs_Stratonovich}. This exercise is left to the reader.
}
\end{remark} 
}

\paragraph{II. Ideal stochastic constraints.} The second way of altering constraints modifies the one-form $ \omega $ so that it includes a dependency on an additional stochastic variable $\mathsf{N} \in \mathbb{R}  ^p$. Let us denote by $\widetilde{\omega} $ this $\mathsf{N}$-dependent one-form on $ Q$ with values in $ \mathbb{R}  ^m $. The explicit construction of $\widetilde{\omega}$ from $ \omega $ depends on the problem at hand.  We associate the following $\mathsf{N}$-dependent linear constraint to the form $ \widetilde{\omega}$: 
\begin{equation}\label{linear_Stoch}
\Delta (q,\mathsf{N}):=\{ v \in T_qQ \mid \widetilde{\omega}(q,\mathsf{N}) \cdot v=0\} \subset T_qQ.
\end{equation} 
The stochasticity is inserted in the dynamics through the constraint as before by assuming that $\mathsf{N}(t)$ verifies the Stratonovich stochastic differential equation
\begin{equation}\label{dN_2}
\mbox{d}\mathsf{N}=F(q,v,\mathsf{N}) \mbox{d}t+ \Sigma  (q,v,\mathsf{N}) \circ \mbox{d}W,
\end{equation} 
for given functions $F, \Sigma : TQ \times \mathbb{R}  ^p \rightarrow  \mathbb{R}  ^p$.
The associated stochastic Lagrange-d'Alembert equations read
\begin{equation}\label{LdA2} 
\left\{
\begin{array}{l}
\vspace{0.2cm}\displaystyle \mbox{d}\frac{\partial L}{\partial v}- \frac{\partial L}{\partial q} \mbox{d}t \in \Delta (q,\mathsf{N}) ^\circ\\
\vspace{0.2cm}\displaystyle \mbox{d}q=v\mbox{d}t\\
\vspace{0.2cm}\displaystyle v \in \Delta (q,\mathsf{N})\\
\mbox{d}\mathsf{N}= F(q,v, \mathsf{N})\mbox{d}t+ \Sigma (q,v, \mathsf{N}) \circ \mbox{d}W,
\end{array}
\right.
\end{equation} 
understood in the Stratonovich sense. In this case, one verifies that $\mbox{d} \left( E(q,v) \right) =0$, where $E(q,v)$ is defined in \eqref{energy_L}, meaning that the energy is preserved along the solutions of \eqref{LdA2}. Note that this setting can be easily generalized to the case when the stochastic process $ \mathsf{N}$ takes values in a manifold. See \cite{FGBPu2016} for details and applications to the rolling ball case. 

\paragraph{Stratonovich form.} Let us assume the existence of local coordinates $(r ^\alpha , s ^a )$ such that $ \widetilde{\omega}$ takes the expression
\begin{equation}\label{local_expression_2} 
\widetilde{\omega}^a (q,\mathsf{N})= ds ^a + \widetilde{A}_ \alpha ^a (r,s,\mathsf{N})dr ^\alpha , \quad a=1,...,m.
\end{equation}
In these coordinates, we can rewrite system \eqref{LdA2} as   
\begin{equation}\label{LdA2_rs} 
\left\{
\begin{array}{l}
\vspace{0.2cm}\displaystyle \mbox{d}\frac{\partial L}{\partial u ^\alpha }- \frac{\partial L}{\partial r ^\alpha }\mbox{d}t= \widetilde{A} _\alpha ^a (r,s,\mathsf{N}) \left(  \mbox{d}\frac{\partial L}{\partial w ^a }- \frac{\partial L}{\partial s ^a } \mbox{d}t\right) \\
\vspace{0.2cm}\displaystyle \mbox{d}r ^\alpha = u ^\alpha \mbox{d}t, \;\; \mbox{d}s ^a =w ^a \mbox{d}t\\
\vspace{0.2cm}\displaystyle  w ^a + \widetilde{A} _\alpha ^a (r,s,\mathsf{N}) u^\alpha =0\\
\mbox{d}\mathsf{N}^j = F ^j (r,s,u,w, \mathsf{N})\mbox{d}t+ \Sigma ^j(r,s,u,w,\mathsf{N}) \circ \mbox{d}W.
\end{array}
\right.
\end{equation} 
Defining the constrained Lagrangian as
\begin{equation}\label{def_LC2} 
\tilde{\mathcal{L}}_C(r ^\alpha ,s ^a ,u ^\alpha  , \mathsf{N} ^j )= L( r ^\alpha ,s ^a ,u ^\alpha ,-\widetilde{A} _\alpha ^a (r,s,\mathsf{N}) u ^\alpha )\,,
\end{equation} 
similar computations as in the deterministic case, yield the first equation of \eqref{LdA2_rs} in the form
\begin{equation}\label{LC_form2} 
\mbox{d}\frac{\partial \tilde{\mathcal{L}}_C}{\partial u ^\alpha }- \frac{\partial \tilde{\mathcal{L}}_C}{\partial r ^\alpha }\mbox{d}t+ \widetilde{A} _\alpha ^a \frac{\partial \tilde{\mathcal{L}}_C}{\partial s ^a }\mbox{d}t=- \frac{\partial L}{\partial w^b}  \left( \widetilde{B} ^b _{ \alpha \beta } w^\beta  \mbox{d}t+\frac{\partial \widetilde{A}  _\alpha ^b }{\partial \mathsf{N} ^j }\mbox{d} \mathsf{N} ^j \right),
\end{equation} 
where $\widetilde{B}(r,s,\mathsf{N})$ is defined in terms of $\widetilde{A}(r,s,\mathsf{N})$ exactly as in \eqref{def_B}. 
The difference with the deterministic counterpart \eqref{constrained_L}, lies in two facts. Firstly, the quantities $A _\alpha ^a $ and $B_{ \alpha \beta } ^a $ of \eqref{constrained_L} are replaced by $\widetilde{A}_\alpha ^a$ and $\widetilde{B}_{ \alpha \beta } ^a$ which now depend on $\mathsf{N}$. As a consequence, the constrained Lagrangian $\tilde{\mathcal{L}}_C$ also depends on $\mathsf{N}$ through its dependence on $\widetilde{A}_ \alpha ^a $ in \eqref{def_LC2}. Secondly, there is the additional last term that involves $\mbox{d}\mathsf{N}$.

As above, by developing the term $ \mbox{d}\frac{\partial \tilde{\mathcal{L}}_C}{\partial u ^\alpha }$ using the Stratonovich chain rule, and assuming that $ \frac{\partial ^2 \tilde{\mathcal{L}}_C}{\partial u ^\alpha \partial u ^\beta }$ is invertible, we can rewrite the system \eqref{LdA2_rs} in the standard Stratonovich form
for the stochastic processes $(r,s,u,\mathsf{N})$ as

\begin{equation}\label{LdA2_rs_Stratonovich} 
\left\{
\begin{array}{l}
\vspace{0.2cm}\displaystyle \mbox{d}r ^\alpha = u ^\alpha \mbox{d}t, \quad  \mbox{d}s ^a =- \widetilde{A} _\alpha ^a (r,s, \mathsf{N}) u^\alpha  \mbox{d}t\\
\vspace{0.2cm}\displaystyle \mbox{d}  u ^\alpha =P(r,s,u, \mathsf{N}) \mbox{d}t - \left( \frac{\partial ^2 \tilde{\mathcal{L}}_C}{\partial u ^\alpha \partial u ^\beta } \right) ^{-1} \left( \frac{\partial ^2 \tilde{\mathcal{L}}_C}{\partial u ^\beta  \partial \mathsf{N} ^j }+ \frac{\partial L}{\partial w ^b } \frac{\partial \widetilde{A}_ \beta  ^b }{\partial \mathsf{N} ^j } \right)  \Sigma _C ^ j \circ \mbox{d}W\\
\vspace{0.2cm}\displaystyle \mbox{d}\mathsf{N} ^j = F ^j _C \mbox{d}t+ \Sigma ^j_C  \circ \mbox{d}W,
\end{array}
\right.
\end{equation} 
where $F_C$ is defined by $F_C(r^\alpha , s^a ,u ^\alpha , \mathsf{N} ^j ):= F ( r ^\alpha ,s ^a ,u ^\alpha ,-\widetilde{A} _\alpha ^a (r,s, \mathsf{N}) u ^\alpha , \mathsf{N} ^j )$, similarly for $\Sigma_C $. The function $P(r,s,u, \mathsf{N})$ has an explicit expression involving the partial derivatives of $\tilde{\mathcal{L}}_C$, the partial derivative $ \frac{\partial L}{\partial w ^b }$, and the quantities $\widetilde{A}_ \alpha ^b $, $\frac{\partial \widetilde{A}_ \alpha ^b}{\partial \mathsf{N} ^j } $, $\widetilde{B}_{ \alpha \beta } ^b $.

\medskip

Expressing the two variants of the stochastic Lagrange-d'Alembert equation in their Stratonovich forms \eqref{LdA1_rs_Stratonovich} and \eqref{LdA2_rs_Stratonovich} helps to identify the role played by the stochastic processes $N$ and $\mathsf{N}$. One  observes that in absence of these noisy perturbations of the constraints, both  \eqref{LdA1_rs_Stratonovich} and \eqref{LdA2_rs_Stratonovich} coincide and recover the deterministic Lagrange-d'Alembert equations. It would be interesting to explore how the diffusion following from equations \eqref{LdA2_rs_Stratonovich} relate to the nonholonomic  diffusion derived in \cite{HoRa2015}, especially in the presence of symmetry. We will postpone the study of this interesting question for future work.

\rem{ 
We shall show below in \S\ref{subsec_Rolling}, that the equations for the rolling ball presented earlier can be obtained as symmetry reduced versions of either \eqref{LdA1} or \eqref{LdA2}. The reduction process being quite involved, we shall first present here a simpler case of reduction, namely, the case of a $G$-invariant system on the Lie group $G$.
} 

\subsection{Stochastic Euler-Poincar\'e-Suslov systems}\label{EPSuslov}

Let us now consider the corresponding two stochastic extensions of the Euler-Poincar\'e-Suslov equations \eqref{EPS}. In the first case, we assume that the functions $F$ and $ \Sigma $ in \eqref{LdA1} are $G$-invariant, so we have $F(g,v,N)=f( g ^{-1} v,N)$ and $ \Sigma (g,v,N)= \sigma (g^{-1} v,N)$, for functions $f, \sigma :\mathfrak{g}  \times \mathbb{R}  ^m \rightarrow \mathbb{R}  ^m $. Using the (left) $G$-invariance of the Lagrangian and the constraints, the stochastic Lagrange-d'Alembert equations \eqref{LdA1} can be rewritten in terms of the reduced quantities as 
\begin{equation}\label{Stoch_EPS1} 
\left\{
\begin{array}{l}
\vspace{0.2cm}\displaystyle \mbox{d} \frac{ \delta  \ell}{ \delta   \xi } - \operatorname{ad}^*_\xi  \frac{ \delta  \ell}{ \delta   \xi }   \mbox{d}t \in (\mathfrak{g}  ^\Delta ) ^\circ, \qquad  \omega (e) \cdot  \xi =N,\\
\mbox{d}N= f( \xi , N)\mbox{d}t+ \sigma ( \xi , N) \circ \mbox{d}W.
\end{array}
\right.
\end{equation} 
For the second case, we assume that the $\mathsf{N}$-dependent one-form $\widetilde{ \omega }$ is $G$-invariant, i.e., $\widetilde{ \omega }(g,\mathsf{N}) \cdot v= \widetilde{ \omega }(e,\mathsf{N}) \cdot g^{-1} v$, so we obtain the $\mathsf{N}$-dependent subspace
\[
\mathfrak{g}  ^\Delta (\mathsf{N}):=\{ \xi \in \mathfrak{g}  \mid \widetilde{ \omega }(e,\mathsf{N}) \cdot\xi =0\} \subset \mathfrak{g}.
\]
The stochastic Lagrange-d'Alembert equations \eqref{LdA2} can be rewritten in terms of the reduced quantities as 
\begin{equation}\label{Stoch_EPS2}
\left\{
\begin{array}{l}
\vspace{0.2cm}\displaystyle \mbox{d} \frac{ \delta  \ell}{ \delta   \xi } - \operatorname{ad}^*_\xi  \frac{ \delta  \ell}{ \delta   \xi }   \mbox{d}t \in (\mathfrak{g} ^\Delta (\mathsf{N})  ) ^\circ, \qquad  \xi \in \mathfrak{g} ^\Delta (\mathsf{N}) ,\\
\mbox{d}\mathsf{N}= \mathsf{f}( \xi , \mathsf{N})\mbox{d}t+ \mathsf{\sigma }( \xi , \mathsf{N}) \circ \mbox{d}W.
\end{array}
\right.
\end{equation}
We shall consider below a more general reduction process that allows us to derive the stochastic equations for the semidirect products, with applications to systems such as the rolling ball. This reduction process, applied to \eqref{LdA1} or \eqref{LdA2}, is a stochastic extension of the nonholonomic reduction process presented in \cite[\S4.3]{GBYo2015} following \cite{Schneider2002}. As in \S\ref{geom_setting}, we will focus on the group $SE(3)$.

\rem{ 
\begin{remark}[Symmetry breaking due to noise]{\rm As in Remark \ref{functions_of_noise}, one can also consider the more general situation in which the noise $N$ is given as a function $N= \gamma (g,n)$. In this case, the $G$-symmetry of the system may be broken by the presence of the noise and one needs to add to the system the reconstruction equation $\mbox{d}g= \xi g\mbox{d}t$.}
\end{remark} 
} 

\subsection{Rolling ball type systems with noisy nonholonomic constraints}\label{subsec_Rolling} 

We shall now present the reduced version of the stochastic Lagrange-d'Alembert equations \eqref{LdA1} and \eqref{LdA2}, in the context of the geometric setting recalled in \S\ref{geom_setting} and for the ``rolling ball type" constraints considered in \eqref{constraint}. We focus here on the case of the motion on the special Euclidean group $SE(3)$, relevant for physical application of the rolling ball. A reader interested in the results written fully in  abstract geometric form should refer to \cite{FGBPu2016}.

We thus fix a Lagrangian $L:TSE(3) \rightarrow \mathbb{R}  $ and a constraint distribution $ \Delta \subset TSE(3)$ that are both $S ^1 \,\circledS\, \mathbb{R}  ^3 $-invariant, and we consider the associated reduced Lagrangian $\ell: \mathfrak{se}(3)  \times S ^2  \rightarrow\mathbb{R}  $ and the associated constraint subspaces $ \mathfrak{se}(3)  ^ \Delta ( \boldsymbol{\Gamma} ) \subset\mathfrak{se}(3)$.

Following our developments in \S\ref{Stoch_LdA}, we consider 
two stochastic deformations of the nonholonomic constraint.

\paragraph{I. Affine stochastic constraint.} We assume that the functions $F, \Sigma :TSE(3) \times\mathbb{R}  ^3  \rightarrow\mathbb{R} ^3  $ are $S ^1 \,\circledS\,\mathbb{R}  ^3 $-invariant, so that they induce the functions $f, \sigma : \mathfrak{se}(3)  \times S ^2 \times\mathbb{R}  ^3  \rightarrow \mathbb{R}  ^3 $. In the context of constraints of ``rolling ball type" \eqref{constraint}, the reduced version of the stochastic deformation considered in \eqref{affine} and \eqref{dN} thus reads
\begin{equation}\label{noisy_affine} 
\mathbf{Y} = \alpha(\boldsymbol{\Gamma} ) \cdot \boldsymbol{\Omega}  +\mathbf{N} ,
\end{equation} 
where the stochastic process $ \mathbf{N}  (t)  \in\mathbb{R}  ^3  $ verifies the Stratonovich stochastic differential equation
\[
\mbox{d} \mathbf{N} =  f ( \boldsymbol{\Omega}  ,\mathbf{Y} ,\boldsymbol{\Gamma} , \mathbf{N} ) \mbox{d}t+ \sigma (\boldsymbol{\Omega}  ,\mathbf{Y} ,\boldsymbol{\Gamma} , \mathbf{N} ) \circ \mbox{d}W.
\]
Using the (left) $G_{a _0}$-invariance of the Lagrangian and the constraints, the stochastic Lagrange-d'Alembert equations \eqref{LdA1} can be rewritten in terms of the reduced quantities as 
\begin{equation}\label{system1} 
\left\{
\begin{array}{l}
\vspace{0.2cm}\displaystyle  \mbox{d} \frac{\delta \ell}{\delta \boldsymbol{\Omega}  }   + \boldsymbol{\Omega} \times \frac{\delta \ell}{\delta\boldsymbol{\Omega}  }  \mbox{d}t+ \mathbf{Y}\times  \frac{\delta \ell}{\delta \mathbf{Y} }  \mbox{d}t+ \alpha( \boldsymbol{\Gamma} )^{T} \cdot  \left( \mbox{d} \frac{\delta \ell}{\delta\mathbf{Y} } +\boldsymbol{\Omega} \times  \frac{\delta \ell}{\delta \mathbf{Y} }  \mbox{d}t\right)   =\frac{\delta\ell}{\delta\boldsymbol{\Gamma} }\times\boldsymbol{\Gamma}  \mbox{d}t\\
\vspace{0.2cm} \displaystyle  \mbox{d}\boldsymbol{\Gamma}  + \boldsymbol{\Omega} \times\boldsymbol{\Gamma}  \mbox{d}t =0, \quad \mathbf{Y} = \alpha(\boldsymbol{\Gamma} ) \cdot \boldsymbol{\Omega}+ \mathbf{N},
\end{array}
\right.
\end{equation}
understood in the Stratonovich sense. The energy, given in terms of the reduced variables as \begin{equation}\label{reduced_energy} 
E( \boldsymbol{\Omega}  , \mathbf{Y} , \boldsymbol{\Gamma} )= \frac{\delta \ell}{\delta \boldsymbol{\Omega}  } \cdot \boldsymbol{\Omega} +\frac{\delta \ell}{\delta \mathbf{Y} } \cdot \mathbf{Y}   - \ell( \boldsymbol{\Omega}  , \mathbf{Y} , \boldsymbol{\Gamma} ),
\end{equation} 
verifies
\begin{equation} 
\label{energy_noncons}
\begin{aligned}
\mbox{d} \left( E( \boldsymbol{\Omega}  , \mathbf{Y} , \boldsymbol{\Gamma} ) \right) &=\left(  \mbox{d} \frac{\delta \ell}{\delta\mathbf{Y} }+  \boldsymbol{\Omega} \times  \frac{\delta \ell}{\delta\mathbf{Y} }\mbox{d}t\right)  \cdot \left(  \mathbf{Y} - \alpha ( \boldsymbol{\Gamma} ) \cdot \boldsymbol{\Omega}\right) \\
& = \left(  \mbox{d} \frac{\delta \ell}{\delta\mathbf{Y} }+  \boldsymbol{\Omega} \times  \frac{\delta \ell}{\delta\mathbf{Y} }\mbox{d}t\right)\cdot  \mathbf{N}
\end{aligned} 
\end{equation} 
and is therefore not conserved in general.

\paragraph{II. Ideal stochastic constraints.} Following the second way of introducing the stochasticity in \S\ref{Stoch_LdA}, we assume that the one-form defining $ \Delta $ includes a dependency on an additional variable $\mathsf{N} \in \mathbb{R}  ^p$. We also assume that this $\mathsf{N}$-dependent one-form is $S ^1\,\circledS\, \mathbb{R} ^3 $-invariant. Similarly, we assume that the functions $F, \Sigma :TSE(3) \times \mathbb{R}  ^p \rightarrow \mathbb{R} ^p $ are $S ^1\,\circledS\, \mathbb{R} ^3 $-invariant, so that they induce the functions $f, \sigma : \mathfrak{se}(3)  \times S ^2  \times \mathbb{R}  ^p \rightarrow \mathbb{R} ^p  $. In the context of constraints of ``rolling ball type" \eqref{constraint}, the reduced version of the stochastic deformation considered in \eqref{linear_Stoch} and \eqref{dN_2} thus reads
\begin{equation}\label{noisy_affine2}
\mathbf{Y} = \widetilde{\alpha}( \boldsymbol{\Gamma} , \mathsf{N}) \cdot\boldsymbol{\Omega} ,
\end{equation} 
where the stochastic process $\mathsf{N} (t)  \in \mathbb{R}  ^p $ verifies the Stratonovich stochastic differential equation
\[
\mbox{d} \mathsf{N}=  f ( \boldsymbol{\Omega}  ,\mathbf{Y} ,\boldsymbol{\Gamma} , \mathsf{N}) \mbox{d}t+ \sigma ( \boldsymbol{\Omega}  , \mathbf{Y} ,  \boldsymbol{\Gamma} ,\mathsf{N}) \circ \mbox{d}W.
\]
Using the (left) $S ^1 \,\circledS\, \mathbb{R}  ^3 $-invariance of the Lagrangian and of the constraints, the stochastic Lagrange-d'Alembert equations \eqref{LdA2} can be rewritten in terms of the reduced quantities as
\begin{equation}\label{system2} 
\left\{
\begin{array}{l}
\vspace{0.2cm}\displaystyle  \mbox{d}\frac{\delta \ell}{\delta \boldsymbol{\Omega}  }   + \boldsymbol{\Omega} \times \frac{\delta \ell}{\delta\boldsymbol{\Omega}  }  \mbox{d}t+ \mathbf{Y}\times  \frac{\delta \ell}{\delta \mathbf{Y} }  \mbox{d}t+ \widetilde{\alpha}( \boldsymbol{\Gamma} , \mathsf{N})^{T} \cdot  \left(  \mbox{d} \frac{\delta \ell}{\delta\mathbf{Y} } +\boldsymbol{\Omega} \times  \frac{\delta \ell}{\delta \mathbf{Y} }  \mbox{d}t\right) =\frac{\delta\ell}{\delta\boldsymbol{\Gamma} }\times\boldsymbol{\Gamma} \mbox{d}t \\
\vspace{0.2cm} \displaystyle  \mbox{d} \boldsymbol{\Gamma}  + \boldsymbol{\Omega} \times\boldsymbol{\Gamma}  \mbox{d}t  =0, \quad \mathbf{Y} = \widetilde{\alpha}(\boldsymbol{\Gamma}, \mathsf{N}) \cdot \boldsymbol{\Omega}+ \mathbf{N},
\end{array}
\right.
\end{equation} 
understood in the Stratonovich sense. 
In this case, the energy $E( \boldsymbol{\Omega}  , \mathbf{Y} , \boldsymbol{\Gamma} ) $ defined in \eqref{reduced_energy} is conserved since we can compute
\[
\mbox{d} \left( E( \boldsymbol{\Omega}  , \mathbf{Y} , \boldsymbol{\Gamma} ) \right) =\left(  \mbox{d} \frac{\delta \ell}{\delta\mathbf{Y} }+  \boldsymbol{\Omega} \times  \frac{\delta \ell}{\delta\mathbf{Y} }\mbox{d}t\right)  \cdot \left(  \mathbf{Y} - \widetilde{\alpha} ( \boldsymbol{\Gamma}, \mathsf{N} ) \cdot \boldsymbol{\Omega}\right) = 0.
\]

\section{Illustration of general theory for the case of Suslov problem}
\label{sec:Suslov}

\label{sec:Sus_prob} 

The Suslov problem, \cite{Suslov1951} (see also  \cite{Kozlov1988,Bloch2003}) considers the motion of a rigid body with a given point (for example, center of mass) fixed in space,  subject to a constraint on the angular velocity. We have introduced the classical Suslov problem earlier in \eqref{Suslov_simple} in the general context of motion on $SO(3)$ group with constraints. We remind the reader that if $\Lambda(t) \in SO(3)$ is the attitude of the body in space, and $\widehat{\bOm} (t) =\Lambda (t) ^T \dot \Lambda(t)  \in \mso(3)$ is the angular velocity, then the nonholonomic constraint is written in terms of the vector $\mathbf{a}$ that is {\em fixed in the body frame}, as 
\[
\mathbf{a} \cdot \boldsymbol{\Omega} =0. 
\]
We extend \eqref{Suslov_simple} by incorporating possible effects of gravity or other external potential and take the Lagrangian of the form
\[
\ell( \boldsymbol{\Omega}, \boldsymbol{\Gamma}  )= \frac{1}{2} \mathbb{I}\boldsymbol{\Omega} \cdot \boldsymbol{\Omega}-U(\boldsymbol{\Gamma} ),
\]
where $ \boldsymbol{\Gamma} = \Lambda  ^{-1} \mathbf{z}$. Then, the Euler-Poincar\'e-Suslov equations \eqref{Suslov_simple} generalize as
\begin{equation} 
\label{Suslov_inhomog} 
\frac{d}{dt} \mathbb{I}  \boldsymbol{\Omega} +\boldsymbol{\Omega}\times  \mathbb{I}  \boldsymbol{\Omega} =\boldsymbol{\Gamma} \times \frac{\partial U}{\partial\boldsymbol{\Gamma}}  + \lambda \mathbf{a}, \quad \frac{d}{dt} \boldsymbol{\Gamma} = \boldsymbol{\Gamma} \times \boldsymbol{\Omega} ,\quad \mathbf{a} \cdot \boldsymbol{\Omega} =0,
\end{equation}
see \cite{Suslov1951}.
The Lagrange multiplier can be solved by taking the scalar product with $ \mathbb{I}  ^{-1} \mathbf{a} $, which yields:
\begin{equation}\label{lambda} 
\lambda = \frac{ \left( \boldsymbol{\Omega}\times  \mathbb{I}  \boldsymbol{\Omega} +\displaystyle \frac{\partial U}{\partial\boldsymbol{\Gamma}} \times \boldsymbol{\Gamma}  \right)  \cdot \mathbb{I} ^{-1} \mathbf{a} }{ \mathbf{a} \cdot\mathbb{I} ^{-1} \mathbf{a} } .
\end{equation} 

Integrals of motion are considered in \cite{Ko1985}. Evidently, the equation always has the three independent integrals:
\[
E= \frac{1}{2} \mathbb{I}  \boldsymbol{\Omega} \cdot \boldsymbol{\Omega} +U( \boldsymbol{\Gamma} ),\quad  \boldsymbol{\Gamma} \cdot \boldsymbol{\Gamma} ,\quad \mathbf{a}\cdot \boldsymbol{\Omega} 
\]

(a) If $U=0$, and $ \mathbf{a} $ is an eigenvector of $ \mathbb{I}  $, then 
\[
\frac{1}{2} \mathbb{I}  \boldsymbol{\Omega} \cdot \mathbb{I}  \boldsymbol{\Omega} 
\]
is an additional integral. 
From now on, we suppose that $ \mathbf{a} $ is an eigenvector of $ \mathbb{I}  $. Without loss of generality let us take $ \mathbf{a} = \mathbf{e} _3 $.

(b) If $U( \boldsymbol{\Gamma} )= \boldsymbol{\chi}  \cdot\boldsymbol{\Gamma} $ with $ \mathbf{a}\cdot \boldsymbol{\chi}  =0$, then
\[
\mathbb{I} \boldsymbol{\Omega}\cdot \boldsymbol{\chi} 
\]
(Kharlamova) is an additional integral.  Indeed, we have $\frac{d}{dt} \mathbb{I} \boldsymbol{\Omega}\cdot \boldsymbol{\chi}= - (\mathbb{I}   \boldsymbol{\Omega} \times \boldsymbol{\Omega} ) \cdot \boldsymbol{\chi} =0$
since $\boldsymbol{\Omega}\cdot \mathbf{a} =0$ and $ \boldsymbol{\chi} \cdot \mathbf{a} =0$, see p.158 in \cite{Ko1985}.

(c) If $U( \boldsymbol{\Gamma} )= \boldsymbol{\chi}  \cdot\boldsymbol{\Gamma} $ with $\boldsymbol{\chi} = \varepsilon \mathbf{a} $ and if $I_1=I_2$ (Lagrange top), then
\[
\mathbb{I}  \boldsymbol{\Omega}\cdot \boldsymbol{\Gamma} 
\]
is an additional integral. Indeed, we have $\frac{d}{dt} \mathbb{I}  \boldsymbol{\Omega}\cdot \boldsymbol{\Gamma} = \lambda \mathbf{a}\cdot\boldsymbol{\Gamma}$.
However, from \eqref{lambda} we see that $ \lambda =0$ since $I_1=I_2$. This corresponds to the fact that the constraint $\boldsymbol{\Omega} \cdot \mathbf{a} $ is always preserved in a Lagrange top (recall $ \mathbf{a} = \mathbf{e} _3 $).

(d) If $U( \boldsymbol{\Gamma} )=  \frac{\varepsilon }{2}\mathbb{I}  \boldsymbol{\Gamma} \cdot \boldsymbol{\Gamma} $, then
\[
\frac{1}{2} \mathbb{I}  \boldsymbol{\Omega}\cdot \mathbb{I}  \boldsymbol{\Omega} - \frac{1}{2} A\boldsymbol{\Gamma}\cdot \boldsymbol{\Gamma} , \quad A: =\varepsilon \mathbb{I} ^{-1}\det\mathbb{I}  
\]
(Clebsch-Tisserand integral) is an additional integral (Theorem 3 in \cite{Ko1985}). This is  a generalization of situation in (a).

\paragraph{Noisy constraint I.} Let us consider the first stochastic extension introduced in \S\ref{I_and_II}. It corresponds to the stochastic constraint
\begin{equation} 
\label{Noisy_constr_1} 
\mathbf{a} \cdot \boldsymbol{\Omega} =N, \quad  \mbox{d}N= f( \boldsymbol{\Omega} ,\boldsymbol{\Gamma} , N)\mbox{d}t+ \sigma (\boldsymbol{\Omega} ,\boldsymbol{\Gamma} , N) \circ \mbox{d}W,
\end{equation} 
for two functions $f, \sigma : \mathbb{R}  ^3 \times\mathbb{R} ^3 \times \mathbb{R} \rightarrow \mathbb{R}  $.
From the stochastic Euler-Poincar\'e-Suslov equations \eqref{Stoch_EPS1}, we get the system
\begin{equation}\label{case_1} 
\left\{
\begin{array}{l}
\displaystyle\vspace{0.2cm}\mbox{d} \mathbb{I}  \boldsymbol{\Omega} +\boldsymbol{\Omega}\times  \mathbb{I}  \boldsymbol{\Omega} \mbox{d}t=\boldsymbol{\Gamma} \times \frac{\partial U}{\partial\boldsymbol{\Gamma}} \mbox{d}t + \lambda \mathbf{a} \mbox{d}t, \quad \mathbf{a} \cdot \boldsymbol{\Omega} =N,\\
\displaystyle\mbox{d}\boldsymbol{\Gamma} = \boldsymbol{\Gamma} \times \boldsymbol{\Omega}\mbox{d}t ,\quad \mbox{d}N=f\mbox{d}t+ \sigma \circ \mbox{d}W,
\end{array}
\right.
\end{equation} 
whose Lagrange multiplier is computed as
\[
\lambda \mbox{d}t= \frac{ \left( \boldsymbol{\Omega}\times  \mathbb{I}  \boldsymbol{\Omega} +\displaystyle \frac{\partial U}{\partial\boldsymbol{\Gamma}} \times \boldsymbol{\Gamma}  \right)  \cdot \mathbb{I} ^{-1} \mathbf{a}\, \mbox{d}t+ \mbox{d}N}{ \mathbf{a} \cdot\mathbb{I} ^{-1} \mathbf{a} } .
\]
If $N$ is a constant number $c$, then the constraint $ \mathbf{a} \cdot \boldsymbol{\Omega} =c$ corresponds to the inhomogeneous Suslov problem, see \cite{Suslov1951}.

By choosing the eigenvector $\mathbf{a}=\mathbf{e}_3$, equations \eqref{case_1} can be written as
\begin{equation}
\left\{
\begin{array}{l}
\vspace{0.1cm}I _1 \mbox{d} \Omega _1 + \left[ (I_3-I_2) \Omega _2 N -(\bGam\times \frac{\partial U}{\partial \boldsymbol{\Gamma} } )\cdot \mathbf{e}_1\right] \mbox{d} t=0  \\
\vspace{0.1cm}I _2 \mbox{d} \Omega _2 +  \left[ (I_1-I_3) \Omega _1 N -(\bGam\times \frac{\partial U}{\partial\boldsymbol{\Gamma} } )\cdot \mathbf{e}_2\right] \mbox{d}t=0\\
\vspace{0.1cm}\mbox{d} \bGam + \bOm \times \bGam \mbox{d}t =\mathbf{0} \\ 
\vspace{0.1cm}\mbox{d} N  = f( \Omega _1 ,\Omega _2 ,\boldsymbol{\Gamma} ,N)  \mbox{d}t+ \sigma ( \Omega _1 ,\Omega _2 \boldsymbol{\Gamma} ,N) \circ  \mbox{d}W
\end{array}
\right.
\label{Suslov_PE}
\end{equation}
and the Lagrange multiplier $\lambda$ can be computed independently as 
\begin{equation} 
\lambda \mbox{d}t= \left( \bOm \times \mathbb{I}  \bOm + \frac{\partial U}{\partial \boldsymbol{\Gamma} } \times  \bGam \right) \cdot \mathbf{e} _3\, \mbox{d}t+ I _3 \mbox{d}N \, . 
\label{lambda_eq} 
\end{equation}

By assuming $U=0$ for simplicity, so that the variable $ \boldsymbol{\Gamma} $ can be discarded, the associated Fokker-Planck equations for the probability density $p( \Omega _1 , \Omega _2, N,t)$ is
\begin{equation} 
\partial _t p=- \partial _{\Omega_1} ( J_1 \Omega _2 Np)-\partial _{\Omega_2} ( J_1 \Omega _1 Np) - \partial _N(fp)+ \frac{1}{2} \partial_N ( \sigma\partial _N( \sigma p)),
\label{FPSuslov_simple}
\end{equation}
where
\[
J _1 = \frac{I_3-I_2}{I_1}, \quad J _2 = \frac{I_1-I_3}{I_2}.  
\]
The Fokker-Planck equations for the general case can be derived similarly.
\medskip

The stochastic extension \eqref{Suslov_PE} does not preserve the integrals mentioned in (a)--(d) if no further assumptions are made. For example for (a), $U=0$ and we have
\[
\frac{1}{2} \mbox{d}(\mathbb{I}  \boldsymbol{\Omega} \cdot \mathbb{I}  \boldsymbol{\Omega} )= \lambda I_3 N\mbox{d}t.
\]
For the Kharlamova integral (b), $U(\boldsymbol{\Gamma} )=\bchi\cdot \bGam$ and we have
\[
\mbox{d}( \mathbb{I}  \boldsymbol{\Omega} \cdot \boldsymbol{\chi} )= -(\mathbb{I}   \boldsymbol{\Omega} \times \boldsymbol{\Omega} ) \cdot \boldsymbol{\chi}\mbox{d}t
\]
which does not vanish in general. Indeed, the conservation of Kharlamova integral hinges on the fact that the result of $\bOm \times \mathbb{I}   \boldsymbol{\Omega} $ points in $\mathbf{e}_3=\mathbf{a}$ direction, and $\bchi \cdot \mathbf{a}=0$ affords vanishing of the appropriate terms. This condition is broken in the case of noisy constraints \eqref{Noisy_constr_1}. We shall see below under which hypothesis on the body this integral of motion is preserved.
 For the integrals (c) and (d) we have, respectively,
\[
\mbox{d} (\mathbb{I}  \boldsymbol{\Omega} \cdot \boldsymbol{\Gamma} )=I_3\Gamma _3\mbox{d}N \quad\text{and}\quad \frac{1}{2}\mbox{d} \left(  \mathbb{I}  \boldsymbol{\Omega}\cdot \mathbb{I}  \boldsymbol{\Omega} -  A\boldsymbol{\Gamma}\cdot \boldsymbol{\Gamma}\right) = \lambda I_3N \mbox{d}t.
\]  
\paragraph{Noisy constraint II.} The energy preserving stochastic constraint introduced in \S\ref{I_and_II} corresponds to
\begin{equation} 
\mathbf{N} \cdot\boldsymbol{\Omega} =0,\quad \mbox{d} \mathbf{N} =  \mathbf{f} ( \boldsymbol{\Omega} ,\boldsymbol{\Gamma} ,  \mathbf{N} )\mbox{d}t+ \boldsymbol{\sigma}  ( \boldsymbol{\Omega} ,\boldsymbol{\Gamma} ,  \mathbf{N} ) \circ \mbox{d}W,
\label{eqN}
\end{equation} 
for two functions $ \mathbf{f} , \boldsymbol{\sigma}  : \mathbb{R}  ^3 \times\mathbb{R} ^3 \times \mathbb{R} ^3 \rightarrow \mathbb{R}  $.
From the stochastic Euler-Poincar\'e-Suslov equations \eqref{Stoch_EPS2} we get the stochastic system
\begin{equation} 
\label{Suslov_homogeneous} 
\left\{
\begin{array}{l}
\displaystyle\vspace{0.2cm}
\mbox{d} \mathbb{I}  \boldsymbol{\Omega} +\boldsymbol{\Omega}\times  \mathbb{I}  \boldsymbol{\Omega} \mbox{d}t=\boldsymbol{\Gamma} \times \frac{\partial U}{\partial\boldsymbol{\Gamma}} \mbox{d}t + \lambda \mathbf{N} \mbox{d}t, \quad \mathbf{N} \cdot \boldsymbol{\Omega} =0,\\
\mbox{d}\boldsymbol{\Gamma} = \boldsymbol{\Gamma} \times \boldsymbol{\Omega}\mbox{d}t,\quad \mbox{d} \mathbf{N} =\mathbf{f} \mbox{d}t+ \boldsymbol{\sigma} \circ \mbox{d}W,
\end{array}
\right.
\end{equation} 
so the Lagrange multiplier is
\begin{equation} 
\label{lambda_homogeneous} 
\lambda  \mbox{d} t = \frac{  \left( \boldsymbol{\Omega}\times  \mathbb{I}  \boldsymbol{\Omega} +\displaystyle \frac{\partial U}{\partial\boldsymbol{\Gamma}} \times \boldsymbol{\Gamma}  \right)  \cdot \mathbb{I} ^{-1} \mathbf{N} \mbox{d} t  \,  -\bOm \cdot \mbox{d} \bN  }{ \mathbf{N} \cdot\mathbb{I} ^{-1} \mathbf{N} } .
\end{equation} 
As opposed to \eqref{case_1}, the stochastic equation \eqref{lambda_homogeneous} preserves the energy $E= \frac{1}{2} \boldsymbol{\Omega}\cdot \mathbb{I} \boldsymbol{\Omega} -U( \boldsymbol{\Gamma} )$.

The stochastic extension \eqref{Suslov_homogeneous} does not preserve the integrals mentioned in (a)--(d) if no further assumptions are made. Indeed, for the integrals (a)--(d), we have, under the same assumptions as before, except when we consider the fixed vector $ \mathbf{a}=\mathbf{e}_3 $, since this vector  is replaced by the process $ \mathbf{N} $:
\begin{equation}\label{integrals}
\begin{aligned} 
{\rm (a)} &\quad \frac{1}{2} \mbox{d}\left( \mathbb{I}  \boldsymbol{\Omega} \cdot \mathbb{I}  \boldsymbol{\Omega} \right) = \lambda \mathbb{I}  \boldsymbol{\Omega} \cdot \mathbf{N} \mbox{d}t\\
{\rm (b)} &\quad \mbox{d} (\mathbb{I}  \boldsymbol{\Omega} \cdot \boldsymbol{\chi} )= -(\mathbb{I}   \boldsymbol{\Omega} \times \boldsymbol{\Omega} ) \cdot \boldsymbol{\chi}\mbox{d}t+ \lambda \mathbf{N}\cdot\boldsymbol{\chi} \mbox{d}t \textcolor{white}{ \frac{1}{2} } \\
{\rm (c)}&\quad \mbox{d}( \mathbb{I}  \boldsymbol{\Omega} \cdot \boldsymbol{\Gamma} )= \lambda  \boldsymbol{\Gamma}\cdot \mathbf{N} \mbox{d}t \textcolor{white}{ \frac{1}{2} } \\
{\rm (d)}&\quad \frac{1}{2} \mbox{d}\left(   \mathbb{I}  \boldsymbol{\Omega} \cdot \mathbb{I}  \boldsymbol{\Omega} -  A\boldsymbol{\Gamma} \cdot \boldsymbol{\Gamma} \right) = \lambda \mathbb{I}  \boldsymbol{\Omega} \cdot \mathbf{N} \mbox{d}t.
\end{aligned} 
\end{equation}
Note that we can longer assume that the vector $\mathbf{N}$ is an eigenvector of $\mathbb{I}$, since it is not stationary. 

\paragraph{Evolution of the Kharlamova integral with stochastic constraints.} 
Let us now analyze the conservation of the Kharlamova integral $\mathbb{I} \bOm \cdot \bchi$.
In general, this integral is not conserved in both settings of the stochastic nonholonomic constraints.

For the first case, i.e., equations \eqref{case_1}, we compute
\[
\mbox{d}( \mathbb{I}  \boldsymbol{\Omega} \cdot \boldsymbol{\chi} )= \chi _1 (I_2-I_3) \Omega _2 N\mbox{d}t+ \chi _2 (I_3-I_1) \Omega _1 N\mbox{d}t,
\]
where we assumed, as in (b), $U ( \boldsymbol{\Gamma} )= \bchi \cdot \bGam$ with $\bchi \cdot \mathbf{a}=0$ for constant vectors $\bchi$ and $\mathbf{a}= \mathbf{e} _3 $. 
It is easy to see that $\mathbb{I} \bOm \cdot \bchi$ is an integral of motion if either one of the following conditions is satisfied: 
\begin{enumerate} 
\item $ \boldsymbol{\chi} = \mathbf{e}_1$ and $I_1=I_3$, 
\item $\boldsymbol{\chi} = \mathbf{e}_2$ and $I_2=I_3$, or 
\item $ \mathbb{I} =  I_0 {\rm Id}_{3 \times 3}$. 
\end{enumerate} 

On the other hand, for the second case, i.e., equations \eqref{Suslov_homogeneous}, if one of the three cases 1., 2., or 3. above is realised, we get the evolution equation for the quantity $\mathbb{I}  \bOm \cdot \bchi$ as 
\begin{equation} 
\label{Suslov_homogeneous2} 
\mathbb{I}   \mbox{d}   \boldsymbol{\Omega} \cdot \bchi =\lambda \mathbf{N} \cdot \bchi \mbox{d}t  \, , 
\end{equation} 
so $ \mathbb{I}\bOm \cdot \bchi$ is only conserved in time if $\mathbf{N} \cdot \bchi=0$ for all times. This can happen, for example, if $\mathbf{N}(0) \cdot \bchi=0$ and the evolution of $\mathbf{N}$ satisfies 
\begin{equation} 
\label{eq_N2} 
d \mathbf{N} = (\mathbf{g} \times \bchi )\mbox{d} t + (\boldsymbol{\eta} \times \bchi)\circ \mbox{d}W,
\end{equation} 
for arbitrary functions $ \mathbf{g} , \boldsymbol{\eta}$ of $( \boldsymbol{\Omega} , \boldsymbol{\Gamma} , \mathbf{N} )$.

\paragraph{Preservation of the other integrals.} From the evolutions equations \eqref{integrals}, we see that for the second type of stochastic constraints, it is always possible to choose the noise such that one of the integral is preserved.

For the integral (c), it suffices to choose $\mathbf{N}(0) \cdot \boldsymbol{\Gamma} (0)=0$ and the evolution of $\mathbf{N}$ given by 
\begin{equation} 
\label{eq_N2_c} 
\mbox{d} \mathbf{N} = (\mathbf{g} \times\boldsymbol{\Gamma}  )\mbox{d} t + (\boldsymbol{\eta} \times\boldsymbol{\Gamma} )\circ \mbox{d}W,
\end{equation} 
for arbitrary functions $ \mathbf{g} , \boldsymbol{\eta}$ of $( \boldsymbol{\Omega} , \boldsymbol{\Gamma} , \mathbf{N} )$.

For the Clebsch-Tisserand integral (d), which contains (a) as a particular case, it suffices to choose $\mathbf{N}(0) \cdot \mathbb{I}  \boldsymbol{\Omega} (0)=0$ and the evolution of $\mathbf{N}$ given by 
\begin{equation} 
\label{eq_N2_c} 
\mbox{d} \mathbf{N} = (\mathbf{g} \times\mathbb{I}  \boldsymbol{\Omega}  )\mbox{d} t + (\boldsymbol{\eta} \times\mathbb{I}  \boldsymbol{\Omega} )\circ \mbox{d}W,
\end{equation} 
for arbitrary functions $ \mathbf{g} , \boldsymbol{\eta}$ of $( \boldsymbol{\Omega} , \boldsymbol{\Gamma} , \mathbf{N} )$.

\paragraph{Analytical solutions and Fokker-Planck equations.} 
Let us simplify the system further and assume that $ \mathbb{I}  =I_0 {\rm Id}_{3 \times 3}$ and $U=0$ in the first case \eqref{case_1}. Then, the equations of motion become simply 
\begin{equation} 
I_0 \mbox{d} \bOm =I_0 \frac{ \mathbf{a} }{ |\mathbf{a}|^2} \mbox{d} N  \quad \Rightarrow \quad \bOm (t) = \bOm_0 + \frac{ \mathbf{a} }{ |\mathbf{a}|^2}  \left( N (t) - N_0 \right)  \, . 
\label{solN} 
\end{equation} 
It is easy to verify that indeed, $\bOm(t) \cdot \mathbf{a}=N(t)$ for all $t$. Using this solution, we can reduce the evolution equation to that of $N$. 

To contrast this with the second case \eqref{Suslov_homogeneous}, let us also assume $\mathbb{I}=I_0 {\rm Id}_{3 \times 3}$ and $U=0$. Then, we get 
\begin{equation} 
I_0 \mbox{d} \bOm = \lambda \bN \mbox{d} t \, , \quad 
\lambda = - I_0 \frac{ \bOm \cdot \mbox{d} \bN}{|\bN|^2} \, , 
\label{dOmega0} 
\end{equation} 
and, by using $\mbox{d} \mathbf{N} =\mathbf{f} \mbox{d}t+ \boldsymbol{\sigma} \circ \mbox{d}W$,
\begin{equation} 
\mbox{d} \bOm = - \bN  \frac{ \bOm \cdot \mathbf{f}}{|\bN|^2}  \mbox{d} t 
-\bN  \frac{ \bOm \cdot \boldsymbol{\sigma}}{|\bN|^2} \circ \mbox{d} W := 
\alpha \bN \mbox{d} t + \beta \bN \circ \mbox{d} W \, . 
\label{dOmega} 
\end{equation} 
No further progress is possible without additional assumptions on $\mathbf{f}$ and $\boldsymbol{\sigma}$ in \eqref{Suslov_homogeneous}. This shows the fundamental difference between these two cases. 

It is also possible to write the equation of evolution for the probability density. 
Assuming that in the simplest case $\mathbf{f}$ and $\boldsymbol{\sigma}$ do not depend on $\bGam$, we obtain the Fokker-Planck equation for the probability density $p(\bOm,\bN,t)$ as 
\begin{equation}
\partial _t p = - {\rm div}_{\bOm} (\alpha \bN p ) - {\rm div}_{\bN}  (\mathbf{f} p ) 
+ 
\frac{1}{2} 
\left( 
{\rm div}_{\bOm}  \beta \bN + {\rm div}_{\bN} \boldsymbol{\sigma} 
\right) 
\left( 
{\rm div}_{\bOm}  \beta \bN p + {\rm div}_{\bN} \boldsymbol{\sigma}  p 
\right) \, , 
\label{FP_linear} 
\end{equation} 
where it is understood that the diffusion terms in the left parentheses of the last term on the right-hand-side apply as an operator to the the right parentheses of the same term.

\section{Conclusion} 

In these notes, we have outlined the procedure for introducing stochasticity into nonholonomic constraints. We have shown that such introduction of noise allows to preserve various integrals of motion. We have also shown that for general nonholonomic systems, the energy is conserved for arbitrary noisy perturbation connecting the velocities, as long as the perturbation preserves the linear structure of the constraints between the velocity components. On the other hand, the energy is in general not conserved in the case of affine, or inhomogeneous, noisy constraints. 

Furthermore, in \cite{FGBPu2016}, we have proved that  the expectation value for the energy can either increase indefinitely, or remain finite, depending on the type of noise used in the equations. These results are derived using analytical studies stemming from the cases of reduced dynamics, such as rolling of a ball along a one-dimensional line.  The choice of noise in constraints preventing infinite energy growth is an interesting question which deserves further studies. 

Another interesting problem is the study of Fokker-Planck equations such as \eqref{FPSuslov_simple} and \eqref{FP_linear} 
 resulting from the Stratonovich SDEs \eqref{LdA1_rs} or \eqref{LdA2_rs}.  It remains to be seen what conditions on the system and noise lead to the resulting Fokker-Planck equation being hypoelliptic, so that the traditional methods of probability theory \cite{Pa2014} apply here. A direct attempt to use \eqref{LdA1_rs} and \eqref{LdA2_rs} to derive Fokker-Planck equations with standard methods gives highly non-intuitive and cumbersome formulas.  It is possible that a more elegant and geometric approach developed by the method of Hamel applied to nonholonomic constraints \cite{ZeLeBl2012,ShBeZeBl2015}, can be useful in approaching this problem. In this method, the velocity coordinates are chosen in such a way that the nonholonomic constraints take a particularly simple form, which may possibly lead to tractable expressions for the diffusion operator in Fokker-Planck equations. We also defer the discussion of this interesting problem for the future. 

\section*{Acknowledgements} 
We acknowledge fruitful and enlightening discussions with Profs. M. Barlow, L. Bates, A. M. Bloch, D. D. Holm, G. Pavliotis, T. S. Ratiu, J. Sniatycki, and D. V. Zenkov. FGB is partially supported by the ANR project GEOMFLUID 14-CE23-0002-01. VP acknowledges  support from NSERC Discovery Grant and the University of Alberta Centennial Fund.


\end{document}